\documentclass[aps,pra, superscriptaddress]{revtex4-2}
\usepackage[utf8]{inputenc}
\usepackage[T1]{fontenc}
\usepackage{graphicx}
\usepackage{amsmath,amssymb}
\usepackage{hyperref}
\usepackage{physics}
\usepackage{amsmath}
\usepackage{siunitx}
\usepackage{bm}
\usepackage{lineno}
\newcommand{\cint}{C_{\mathrm{int}}}
\newcommand{\cintdirect}{C_{\mathrm{int}}^0}
\newcommand{\cintdirecttarget}{C_{\mathrm{int}}^{0,T}}
\newcommand{\cintdirectretopt}{C_{\mathrm{int}}^{0,\mathrm{opt.}}}
\newcommand{\cintdirectpc}{C_{\mathrm{int}}^{0,\mathrm{PC}}}
\newcommand{\cintdirectcclimit}{C_{\mathrm{int}}^{0,\mathrm{CC}}}
\newcommand{\kint}{\kappa_{\mathrm{int}}}

\newcommand{\lint}{\mathcal{L}_{\mathrm{int}}}
\newcommand{\lclip}{\mathcal{L}_{\mathrm{clip}}}
\newcommand{\lnotclip}{\mathcal{L}_{\mathrm{abs}}}

\newcommand{\woconfocal}{w_{0}^{\mathrm{conf}}}

\newcommand{\idealpc}[1]{#1^{\mathrm{PC}}_{\mathrm{opt. sph.}}}
\newcommand{\idealpcr}{\idealpc{R}}
\newcommand{\inverseeffectivearea}{A^{-1}_{\mathrm{eff}}}
\newcommand{\inverseeffectiveareaemitter}{\inverseeffectivearea(\bm{r_e})}

\begin{document}

\title{Advanced mirror shapes for mode enhancement in plano-concave cavities}
\author{William J. Hughes}
\email{w.j.hughes@soton.ac.uk}
\author{Peter Horak}
\affiliation{Optoelectronics Research Centre, University of Southampton, Southampton, S017 1BJ, UK}

\date{\today}


\begin{abstract}
Optical cavities are frequently used in quantum technologies to enhance light matter interactions, with applications including single photon generation and entanglement of distant emitters. The Fabry-P\'{e}rot resonator is a popular choice for its high optical access and large emitter-mirror separation. A typical configuration, particularly for emitters that should not be placed close to the mirror surface like trapped ions and Rydberg atoms, features two spherical mirrors placed around a central emitter, but this arrangement can put demanding requirements on the mirror alignment. In contrast, plano-concave cavities are tolerant to mirror misalignment and only require the manufacture of one curved mirror, but have limited ability to focus light in the centre of the cavity. Here we show how mirror shaping can overcome this limitation of plano-concave cavities while preserving the key advantages. We demonstrate through numerical simulations that simple mirror shaping can increase coupling between a plano-concave cavity and a central emitter by an order of magnitude, even rivalling misalignment-sensitive concave-concave counterparts for achievable interaction strength. We use these observations to establish the conditions under which plano-concave cavities with shaped mirrors could improve the performance and practicality of emitter-cavity systems.
\end{abstract}

\maketitle

\section*{Introduction}

Fabry-P\'{e}rot cavities are increasingly employed across quantum technology to improve processes such as single photon production~\cite{Law:97, keller:04} and the readout of qubit states~\cite{Bochmann:10, Grinkemeyer:25, Wang:25}, finding application in quantum networking~\cite{Reiserer:15, Ritter:12, Krutyanskiy:23b} and modular quantum computation~\cite{Monroe:14, Li:24, Sinclair:25}. For many applications, high cooperativity, which means a large ratio of coherent emitter-cavity coupling to incoherent decay processes, is crucial. One can modify the Fabry-P\'{e}rot cavity to increase cooperativity by either decreasing the cavity round trip loss or tightening the mode waist at the position of the emitter. The former is difficult because the lower bound to round trip loss is set by technical limits to the mirror surface roughness and the absorption of the mirror material~\cite{Rempe:92, Shadmany:25, Maier:25}. The latter requires either a shorter cavity length, or moving towards a `concentric' geometry in which both mirrors focus directly on the emitter~\cite{Shadmany:25}. However, a very short cavity length is not compatible with emitters that are sensitive to electric field noise near the high-reflectivity dielectric mirrors~\cite{Teller:21}, such as Rydberg atoms~\cite{Davtyan:18, Chen:22} or ions~\cite{Ong:20}, for which the mirrors additionally disrupt the trapping potential~\cite{Podoliak:16, Kassa:25}, and a highly concentric geometry is extremely sensitive to mirror misalignment~\cite{Nguyen:18}. Concave-concave Fabry-P\'{e}rot optical cavities with spherical mirrors thus have a limited ability to achieve high cooperativity, reasonable emitter-mirror separation, and good alignment tolerance simultaneously. To circumvent these limits, alternative designs have featured additional mirrors~\cite{Cox:18, Chen:22} or intra-cavity optics~\cite{Shadmany:25}, but these introduce experimental complexity, include lossy elements in the cavity round trip, or lengthen the round trip time.

In contrast, plano-concave cavities, which have one planar and one curved mirror, do not introduce these disadvantages. They also boast zero sensitivity to the transverse misalignment of the mirrors and require the manufacture of only one curved surface. Plano-concave geometries are therefore often used~\cite{Wang:17, Malmir:22}, particularly for cavities with short lengths, or when the emitter can be placed close to the planar mirror~\cite{Farrow:23}. However, they have limited ability to focus light in the centre of the cavity, which reduces the possible cooperativity in cavities with longer lengths.

Here we show how to overcome this focussing limit with order of magnitude-level improvements. Our approach uses the ability of modern micromirror manufacture techniques, such as focussed ion beam milling~\cite{Dolan:10} or laser ablation~\cite{Hunger:10}, to control the mirror profile shape~\cite{Trichet:15, Walker:21, Ott:16} while achieving low loss~\cite{Maier:25, Gao:25}. While earlier work has considered the benefits of mirror shape control in concave-concave geometries~\cite{Ferdous:14, Podoliak:17, Karpov:22_1, Karpov:22_2, Karpov:23}, we instead investigate plano-concave cavities and show how they can provide optical performance competitive with spherical concave-concave cavities while maintaining their advantages of mode stability and alignment insensitivity. To do this, we first define our metric for evaluating cavity performance, before finding the limits to this metric in concave-concave and plano-concave cavities with spherical mirrors. We then discuss possible profiles for shaped mirrors, determine the resultant performance in plano-concave designs, and discuss the sensitivity of shaped designs to manufacturing error. We finally summarise the conditions under which plano-concave cavities with shaped mirrors can offer substantial advantages in experimental design.

\section*{Emitter-independent cavity metric}
A common metric for emitter cavity systems, generally desired to be large, is the cooperativity
\begin{equation}
C = \frac{g^2}{2\kappa \gamma},
\label{eq: cooperativity rates expression}
\end{equation}
which represents the strength of coherent photon-emitter coupling compared to incoherent decay processes, where $g$ is the rate of coherent interaction between the emitter and the single cavity mode, $\kappa$ is the cavity amplitude decay rate, and $\gamma$ is the amplitude decay rate of the emitter's excited state, which accounts for spontaneous emission. However, $C$ is often a flawed \textit{single} metric because many applications, such as single photon generation, require not just high cooperativity, but high transmitted fraction (in this example to generate the photon within the cavity and then extract it respectively). In general, particular applications will function best with an optimal transmission that compromises between cooperativity (reduced by increased transmission) and transmitted fraction (increased with transmission). This optimal output transmission can, within a certain tolerance, generally be manufactured by adjusting the mirror coating.

This reasoning leads to a more universal metric known as the internal cooperativity,
\begin{equation}
\cint = \frac{g^2}{2\kint \gamma},
\label{eq: internal cooperativity rates expression}
\end{equation}
where $\kint$ is the partial cavity amplitude decay rate due to internal (i.e. non-transmissive) losses ($\lint$). The internal cooperativity quantifies the innate potential of the cavity geometry (i.e. without deliberate coupling to the outside world) and has the key advantage that it is not dependent on the controllable transmission. The internal cooperativity serves as the single figure of merit for many cavity applications, including photon generation~\cite{Goto:19, Utsugi:22, Hughes:24_2}, adiabatic intracavity gate protocols~\cite{Goto:10}, and intracavity cat-state generation~\cite{Kikura:25}, on the understanding that the output mirror transmission will be optimised for the desired application.

The internal cooperativity has an alternative form that highlights the focussing of the mode on the emitter, $\inverseeffectivearea(\bm{r_e})$, which is the ratio of the intensity of the forward-running cavity mode at the emitter (positioned at $\bm{r_e}$) to the total forward-running power~\cite{Hughes:25},
\begin{equation}
\cint = \alpha_B \cintdirect, \quad \cintdirect = \frac{3 \lambda^2 \inverseeffectivearea(\bm{r_e})}{\pi \lint}, 
\label{eq: unit branching internal cooperativity equation}
\end{equation}
where $\alpha_B$ is the branching ratio of the excited state emission to the cavity mode, which accounts for the spontaneous emission branching ratio and the polarisation power overlap to the desired cavity mode, $\lambda$ is the emitter transition wavelength, and it is assumed that the emitter is moved perfectly to an antinode of the intra-cavity standing wave formed by superposition of the forward-running and backward-running modes. Equation~(\ref{eq: unit branching internal cooperativity equation}) introduces $\cintdirect$, which is proportional to $\cint$. Because we can calculate $\cintdirect$ without specifying the emitter, we quote its value throughout the manuscript, but an optimisation of $\cintdirect$ implicitly optimises the performance-relevant $\cint$.

For cavities with spherical mirrors, the eigenmodes are a family of Gaussian beams with central waist $w_0$ in a transverse plane determined by the cavity geometry~\cite{Kogelnik:66}. Of this family, it is most common to use the `fundamental' mode due to its more compact and simpler spatial structure. An emitter coupling to the  transverse centre of the fundamental mode at an axial position where the waist is $w_e$ has an inverse effective area and, from Eq.~(\ref{eq: unit branching internal cooperativity equation}), an internal cooperativity of  
\begin{equation}
    \inverseeffectivearea = \frac{2}{\pi} \frac{1}{w_e^2}, \quad \cintdirect = \frac{6 \lambda^2}{\pi^2 w_e^2 \lint}.
\label{eq: gaussian waist internal cooperativity}
\end{equation}

\section*{Concave-concave and plano-concave cavities with spherical mirrors}
To assess whether non-spherical mirrors can improve $\cintdirect$, we must first establish the limits to $\cintdirect$ in concave-concave and plano-concave cavities with spherical mirrors. For each case we assume that all aspects of the cavity except for the mirror profile have been determined beforehand. These are $L$, the length of the cavity, which may be restricted by the experimental geometry or minimum emitter-mirror distance, $D$, the mirror diameter, which may be restricted by the maximum machinable surface area or optical access constraints, $\lambda$, the resonant wavelength set by the emitter transition, and $\lnotclip$, the lower limit to internal loss containing scattering and absorptive, but not clipping, losses, set by the mirror coating and small scale surface roughness~\cite{Bennett:92}.

The typical configuration of the concave-concave resonator has two mirrors of identical curvature $R$~\cite{Takahashi:20, Krutyanskiy:23a}. The smallest waist of the mode ($w_0$) is located in the centre of the cavity, and is therefore also the waist at the emitter $w_e$, with
\begin{equation}
    w_e = w_0 = \woconfocal\left(\frac{2R}{L} -1\right)^{\frac{1}{4}}, \quad
    \woconfocal = \sqrt{\frac{\lambda L }{2\pi}},
\end{equation}
where $\woconfocal$ is the central waist of a `confocal' cavity where $R$ is set to the same value as $L$. The waist at the mirror is

\begin{equation}
    w_m = \woconfocal \sqrt{\left(\frac{w_0}{\woconfocal}\right)^2 + \left(\frac{\woconfocal}{w_0}\right)^2},
    \label{eq: waist at mirror}
\end{equation}
which has a minimum value $\sqrt{2}\woconfocal$, satisfied in the confocal configuration.

The internal cooperativity depends upon the cavity mode through the ratio $\inverseeffectiveareaemitter/\lint$ (see Eq.~(\ref{eq: unit branching internal cooperativity equation})). The numerator of this ratio is the tightness of the mode focus on the emitter, but, to understand the impact of the cavity mode on the denominator, we may decompose the internal loss into
\begin{equation}
        \lint = \lclip + \lnotclip,
\end{equation}
where $\lclip$ is loss from the mode escaping beyond the edge of the mirrors. Decreasing the radius of curvature $R$ of the cavity mirror from $R=L$ (confocal limit) towards $R=L/2$ (concentric limit) reduces $w_0$ and thus increases $\inverseeffectiveareaemitter$. However, $w_m$ will increase and thus $\lclip$ will increase dramatically. The optimum mirror profile is found when $R$ is reduced until $\lclip$ becomes a non-negligible fraction of $\lint$. We can write this condition as $\lclip=\chi\lnotclip$, where $\chi$ is a small fraction of unity, but not orders of magnitude smaller than unity~\cite{Gao:23}. Assuming for algebraic convenience that the mirror diameter $D$ is sufficiently large that the central waist is much smaller than $\woconfocal$, we get a maximised $\cintdirect$ of (see Methods)
 \begin{equation}
    \cintdirectcclimit = \frac{1}{\left\{-\frac{1}{2}\ln\left({\frac{\chi}{2}\lnotclip}\right)\right\}}\frac{6 D^2}{L^2  \lnotclip},
\label{eq: cint concave concave large diameter limit}
\end{equation}
where the precise value of $\chi$ has very little impact, so need not be carefully determined. The right-hand side quotient highlights the importance of the cavity numerical aperture $D/L$.

In many actual systems, the limit (Eq.~(\ref{eq: cint concave concave large diameter limit})) cannot be reached because, if the mirrors are not perfectly transversely aligned, the mode will tilt at an angle, determined through geometric cavity theory~\cite{Siegman:86, Hughes:23},
\begin{equation}
    \theta = -\frac{M}{L}\left(\frac{\woconfocal}{w_0}\right)^4,
\end{equation}
where $M$ is the transverse misalignment between the mirrors. The mode angle diverges very strongly as $w_0$ decreases to increase $\cint$, dramatically increasing clipping loss~\cite{Gao:23}. This means that often the alignment tolerance, rather than the cavity numerical aperture, limits the achievable $\cintdirect$ in concave-concave systems.

The plano-concave geometry is naturally insensitive to the transverse mirror misalignment, provided that the mode does not clip outside the edge of the planar mirror. However, when the non-planar mirror is spherical, the cavity modes focus on the planar mirror, not the emitter in the centre of the cavity. For a given $L$, an elementary calculation finds the smallest possible waist at the emitter for mirror radius $R=\idealpcr=5L/4$ taking the value
\begin{equation}
    \idealpc{(w_e)} = \sqrt{\frac{\lambda L}{\pi}} = \sqrt{2}\woconfocal.
\end{equation}
The waist of this mode at the mirror ($w_m = \sqrt{5}\woconfocal$) does not depend upon mirror diameter. In the large mirror diameter limit ($D \gg 2w_m$), the clipping losses are negligible, and, using Eq.~(\ref{eq: gaussian waist internal cooperativity}), the $\cintdirect$ of a plano-concave cavity of length $L$ is limited to 
\begin{equation}
    \cintdirectpc = \frac{6 \lambda}{\pi L \lnotclip}.
    \label{eq: cint pc limit}
\end{equation}
For future reference, we define $\idealpc{\epsilon}$ as the fundamental Gaussian mode of this plano-concave cavity whose spherical mirror has $R=\idealpcr$.

\section*{Suggested surface designs}
For plano-concave cavities, restricting the non-planar mirror to a spherical shape limits $\cintdirect$ (Eq.~(\ref{eq: cint pc limit})) far below the concave-concave limit (Eq.~(\ref{eq: cint concave concave large diameter limit})) set by the cavity solid angle. We will demonstrate that shaping the non-planar mirror can bridge the gap between the two limits, and thus significantly increase $\cintdirect$ for plano-concave cavities. In order to calculate the modes of these plano-concave cavities with shaped mirrors, we use the mode mixing method~\cite{Kleckner:10}, which has been widely used to model cavities with non-spherical mirror profiles~\cite{Benedikter:15, Podoliak:17, Walker:21, Karpov:22_1, Karpov:22_2, Karpov:23, Hughes:23}.

We will use two approaches to design surfaces, which we restrict to having zero azimuthal dependence. The first is the `retroreflective optimisation method'~\cite{Hughes:25}, in which a target cavity mode is optimised for $\cintdirect$, before the mirror surface profile is constructed to retroreflect this target mode. This method is useful because it approximately finds the limits of $\cintdirect$ achievable with arbitrary mirror shaping.

The second approach is to choose non-spherical mirror shapes with a few variable parameters. We will study three such shapes in this manuscript: The Gaussian-shaped mirror, the `dual curvature mirror', and the `spline mirror'. 

We consider Gaussian surfaces because laser ablation naturally makes mirrors more Gaussian-shaped than spherical~\cite{Muller:10}, and because they can improve emitter coupling in concave-concave cavities~\cite{Podoliak:17}. The surface profile is
\begin{equation}
    z(r) = D_G\left\{1-\exp\left(-\frac{r^2}{w_G^2}\right)\right\} , \quad
    D_G = \frac{w_G^2}{2 R},
    \label{eq: Gaussian surface equation}
\end{equation}
where $R$ is the central radius of curvature of the mirror, $w_G$ is the 1/e-waist of the Gaussian profile, the derived quantity $D_G$ is the depth of the central depression from the asymptotic flats, and $r$ is the radial coordinate of the rotationally-symmetric surface profile.

To increase mirror design flexibility, we also study a `dual curvature' profile that transitions smoothly from a central curvature $R_c$ in the middle of the mirror to an outer curvature $R_o$ at the edge. The gradient of this profile is
\begin{equation}
    \frac{dz}{dr} = \left(1 - \sigma(r)\right) \frac{x}{R_c} + \sigma(r)\frac{x}{R_o}, \quad \sigma(r)  = \frac{1}{2}\left( 1 + \frac{\tilde{r}}{\sqrt{1 + \tilde{r}^2}}\right), \quad
        \tilde{r}  = \frac{r - \delta_\sigma}{w_\sigma},
    \label{eq: dual curvature profile gradient}
\end{equation}
where $\tilde{r}$ is a scaled radial coordinate, and $\sigma(r)$ is a `sigmoid' function that transitions between zero and unity. The sigmoid parameter $\delta_\sigma$ sets the radial centre of the transition between $R_c$ and $R_o$, and $w_\sigma$ sets the radial width of the gradient transition. This yields an algebraic profile with four surface parameters
\begin{equation}
\begin{aligned}
    z(r)  =  \frac{1}{2}\left(\frac{x^2}{2R_c} + \frac{x^2}{2R_o}\right) & + \\
    \frac{1}{2}\left(\frac{1}{R_o} - \frac{1}{R_c}\right) \, \, & \Bigg\{\frac{1}{2}w_\sigma^2 \tilde{r}\sqrt{1 + \tilde{r}^2} + w_\sigma \delta_\sigma \sqrt{1 + \tilde{r}^2} \\ &  - \frac{1}{2}w_\sigma \delta_\sigma \sqrt{1 + \left(\frac{\delta_\sigma}{w_\sigma}\right)^2} - \frac{1}{2}w_\sigma^2 \left[\mathrm{arsinh}(\tilde{r}) - \mathrm{arsinh}\left(\frac{-\delta_\sigma}{w_\sigma}\right)\right] \Bigg\}. 
\end{aligned}
\label{eq: Dual curvature surface equation}
\end{equation}

Finally, to encode a surface in a flexible number of free parameters, we parametrise the mirror surface through a parabolic curvature $R$ and an additional cubic spline. The surface amplitude follows
\begin{equation}
    z(r) = \frac{r^2}{2R} + s(r),
    \label{eq: spline surface equation}
\end{equation}
where $s(r)$ is a cubic spline distortion from parabolic passing through chosen points. The spline's $N_p + 1$ points are equispaced between radial coordinates of $0$ and $D/2$ inclusive, where the first point (at the origin) is always set to zero, and the remaining $N_p$ points (along with $R$) are free parameters. The spline passes smoothly through the points, with zero gradient enforced at the origin.

\section*{Performance improvements with shaped mirrors}

To assess the potential of these mirror shapes to improve cavity performance, we simulate their use in an example cavity geometry. This geometry has length \SI{1}{\milli\metre}, resonant wavelength \SI{1033}{\nano\metre} to target an example dipole transition in $\mathrm{Sr}^{+}$ ions, mirror diameter \SI{300}{\micro\metre}, and $\lnotclip$ 20ppm. The length is chosen to avoid the strong impacts of dielectric mirrors on trapped ions seen at lengths of a few hundreds of microns~\cite{Teller:21, Kassa:25}. Details of the numerical parameters and the particular methods used for the simulations in this manuscript are found in Methods.

\begin{figure}[ht!]
\centering\includegraphics[width=16.5cm]{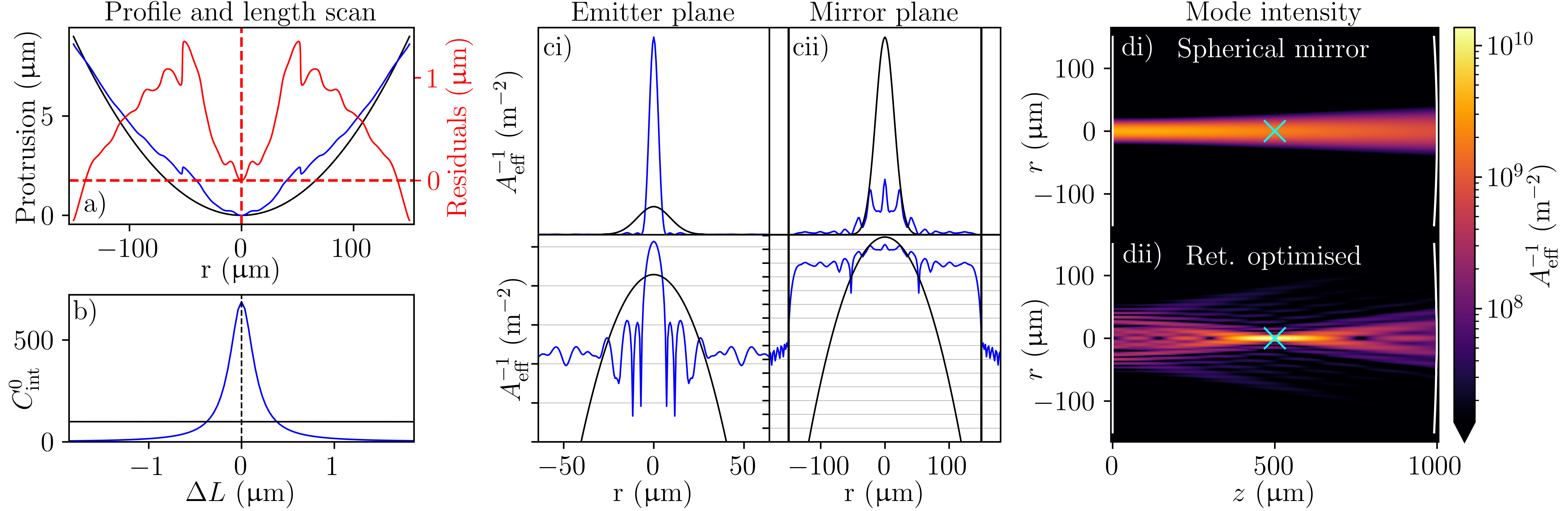}
\caption{Results of retroreflective optimisation for a cavity with the example geometry ($L=$~\SI{1}{\milli\metre}, $\lambda=$~\SI{1033}{\nano\metre}, $D=$~\SI{300}{\micro\metre}, $\lnotclip=$~20ppm). a) Surface profile of the (black) best spherical mirror and (blue) retroreflective optimised mirror with (red) residuals overlaid. b) Performance of a plano-concave cavity with the optimised non-planar mirror as a function of length change from the nominal value. The black line marks $\cintdirectpc$ over the length scan. c) The running-wave intensity (linear scale on top, log scale bottom) of the (blue) cavity eigenmode and (black) best spherical mode $\idealpc{\epsilon}$ in the i) emitter and ii) mirror planes respectively. The vertical lines in ii) represent the edge of the mirror. d) Mode intensity in the $xz$-plane ($y=0$) for i) $\idealpc{\epsilon}$ and ii) the optimised mode. The planar mirror is at $z=0$ (left) and the non-planar mirror at $z=L$ (right). The cyan cross marks the emitter position.}
\label{fig: retroreflective_optimisation_example}
\end{figure}

For retroreflective optimisation (Fig.~\ref{fig: retroreflective_optimisation_example}), a complicated surface design featuring a tight central curvature and small scale details (Fig.~\ref{fig: retroreflective_optimisation_example}a) increases $\cintdirect$ far above $\cintdirectpc$, but only over a very limited length range (Fig.~\ref{fig: retroreflective_optimisation_example}b). This superior performance derives from focussing much more tightly on the emitter with a specific mode that fits exactly into the mirror's finite diameter (Fig.~\ref{fig: retroreflective_optimisation_example}c-d). This confirms the principle that mirror shaping can exploit the full extent of the non-planar mirror to circumvent the performance limitations brought by Gaussian eigenmodes. However, the complicated profile would be challenging to fabricate, and the extreme length sensitivity, down to even a wavelength scale, means that the cavity could not even be freely tuned onto a desired resonant frequency without affecting $\cint$ considerably.

\begin{figure}[ht!]
\centering\includegraphics[width=16cm]{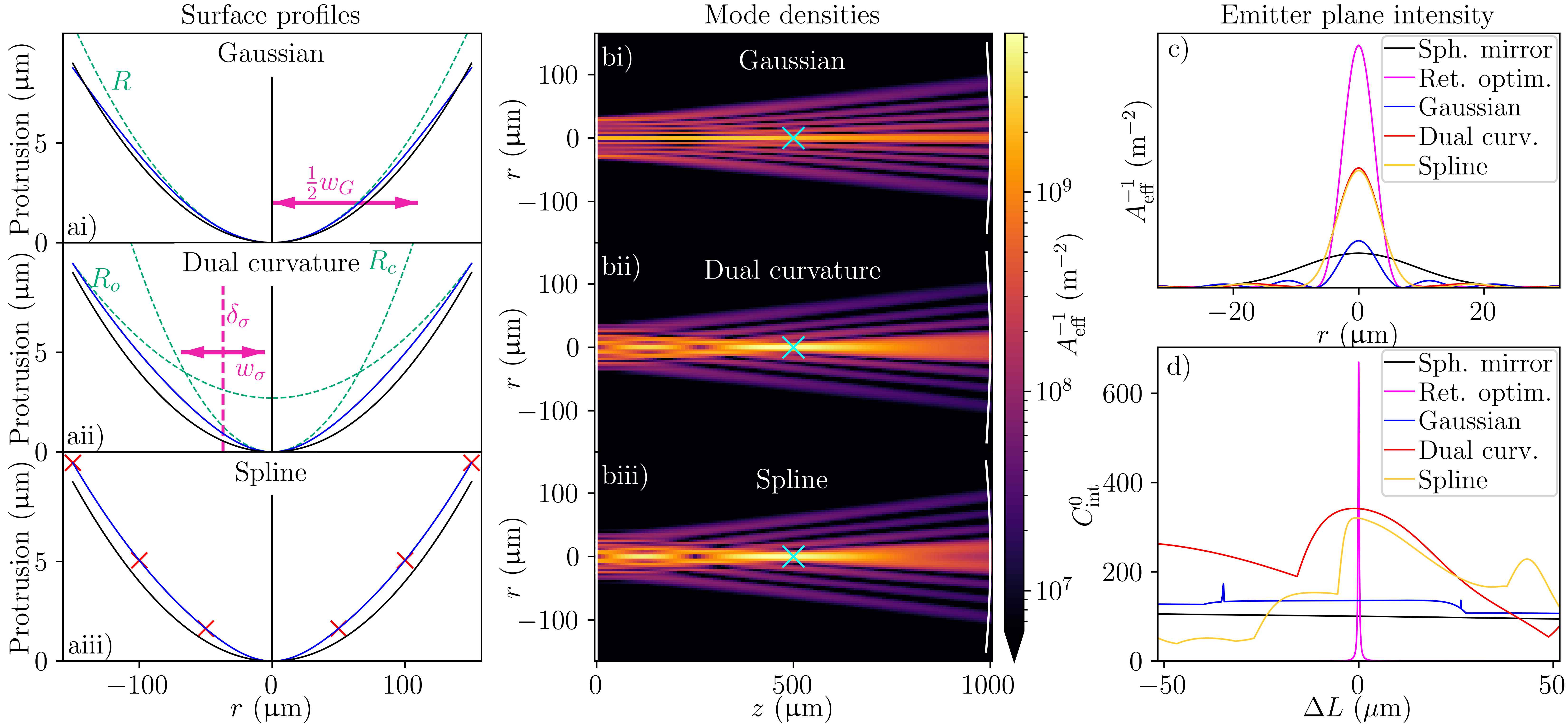}
\caption{Example mirror profiles and resulting high-performing cavities for the example geometry ($L=$~\SI{1}{\milli\metre}, $\lambda=$~\SI{1033}{\nano\metre}, $D=$~\SI{300}{\micro\metre}, $\lnotclip=$~20ppm). a) Example mirror shapes with i) Gaussian, ii) dual-curvature, and iii) 3-point spline designs (blue) compared to the best spherical mirror (black). In ai) annotations mark the central radius $R$ and waist $w_G$ of the Gaussian profile (labelled as $\frac{1}{2}w_G$ because the Gaussian waist exceeds half the mirror diameter). In aii) annotations mark the four parameters of the dual curvature mirror (central $R_c$ and outer $R_o$ radii of curvature, and transition radius $\delta_\sigma$ and width $w_\sigma$). In aiii) the crosses mark the coordinates of the chosen spline points. b) Mode intensities in the $xz$-plane ($y=0$) for cavities with the Gaussian, dual curvature, and 3-point spline mirrors shown in a), where cyan crosses mark the emitter positions. c) Mode intensity in the emitter plane and d) $\cintdirect$ as a function of length change from the nominal value for the surface profiles in a), and for the optimised spherical mirror and the retroreflective optimised mirror shown in Fig.~\ref{fig: retroreflective_optimisation_example}.}
\label{fig: designed surfaces}
\end{figure}

For the designed surface shapes of Eqs.~(\ref{eq: Gaussian surface equation}), (\ref{eq: Dual curvature surface equation}), and (\ref{eq: spline surface equation}) (see Fig.~\ref{fig: designed surfaces}), we also find surfaces that outperform all spherical mirrors. Again, high-performing mirror shapes feature a tighter central curvature than the best spherical mirror (Fig.~\ref{fig: designed surfaces}a), and the modes exploit the space on the non-planar mirror to focus strongly on the emitter, albeit to a lesser extent than for the retroreflective optimised mirror (Fig.~\ref{fig: designed surfaces}b-c). The length sensitivity of the few-parameter mirror surfaces depends upon the mirror shape, but is generally much reduced (Fig.~\ref{fig: designed surfaces}d), meaning the length may be scanned to select resonances over many free spectral ranges without an appreciable drop in $\cintdirect$. Note that the mirrors surfaces depicted in Fig.~\ref{fig: designed surfaces} are not global maxima, but instead chosen by hand for their combination of near-optimal performance and robustness to parameter change. This represents the likely case that an experimental design would compromise slightly on performance for parameter robustness (further discussion in a later section of the manuscript).

\begin{figure}[ht!]
\centering\includegraphics[width=11cm]{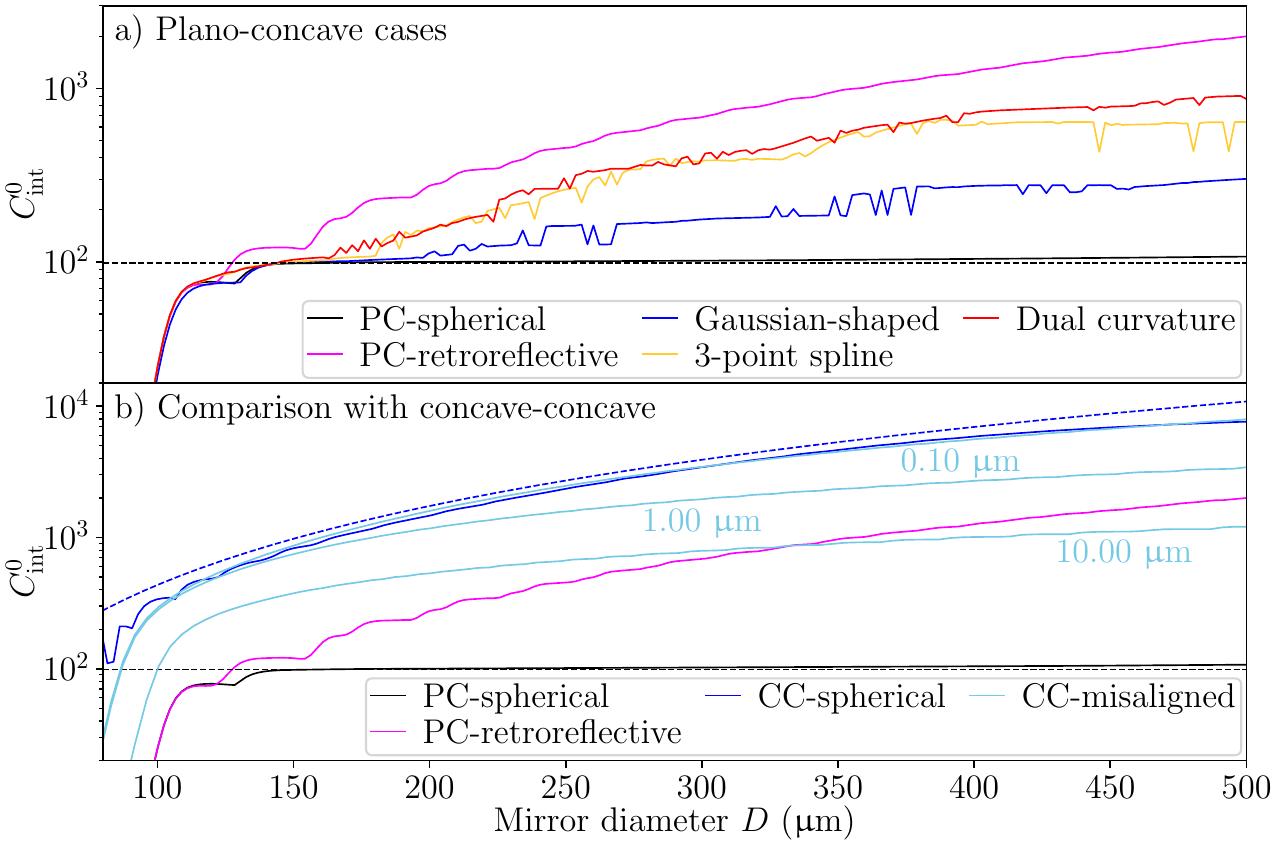}
\caption{Achievable $\cintdirect$ as a function of mirror diameter for different surface profile designs. The cavities have $\lambda=$~\SI{1033}{\nano\metre}, $L=$~\SI{1}{\milli\metre}, and $\lnotclip=$~20ppm. a) Comparison of mirror shaping in plano-concave cavities. The different lines represent shaping strategies labelled in the legend, where `PC' is shorthand for plano-concave. The black, dashed, perfectly horizontal line marks $\cintdirectpc$ (Eq.~(\ref{eq: cint pc limit})). b) Comparison of plano-concave designs to concave-concave designs, where `CC' in the legend denotes `concave-concave'. For the concave-concave cavities with transverse misalignment, the misalignment is labelled on the corresponding line. The blue dashed line graphs $\cintdirectcclimit$ (Eq.~(\ref{eq: cint concave concave large diameter limit})).}
\label{fig: master scan combined}
\end{figure}

Having established that mirror shaping can improve $\cintdirect$ in plano-concave designs, we now assess how the achievable improvements vary with geometric constraints by optimising mirror profiles across a scan of the mirror diameter. To discourage the optimisation from finding few-parameter mirror designs (Gaussian, dual-curvature, and spline) with the extreme length sensitivity of retroreflective optimised designs, we simulate each mirror at the nominal length $L$, and $L \pm \lambda/4$, returning the lowest $\cintdirect$ to the optimisation routine. For spherical mirrors (Fig.~\ref{fig: master scan combined}a black line), we see that the performance is limited to $\cintdirectpc$ for mirror diameters large enough to contain $\idealpc{\epsilon}$ without significant clipping losses. If mirror diameter is decreased, the mirror curvature will first decrease to reduce the mode size on the non-planar mirror, with $\cintdirect$ plateauing again at the minimum mirror waist size for $R=L$ where degenerate mode mixing further protects against clipping loss~\cite{Hughes:23}. If the mirror diameter is decreased further, the cavity cannot sustain any low loss modes, and $\cintdirect$ reduces quickly. The numerical performance also slightly exceeds $\cintdirectpc$ for large mirror diameters. This is because these cavities support many Laguerre-Gauss modes with negligible loss which, in the absence of mode mixing, would all have $\cintdirect = \cintdirectpc$. However, slight clipping can deform the eigenmodes to marginally increase the intensity of one mode (typically a higher order mode due to stronger mixing) on the emitter. Despite the mixing of higher order modes, the fundamental mode of the cavity will have a $\cintdirect$ very close to $\cintdirectpc$ for sufficiently large mirror diameters.

The other lines in Fig.~\ref{fig: master scan combined}a, show that mirror shaping can strongly increase performance if there is unused space on the non-planar mirror (indicated by the spherical mirror performance hitting $\cintdirectpc$), even to the order of magnitude level for large diameters. Though the few-parameter mirrors cannot reach the $\cintdirect$ of the retroreflective-optimised mirror, the dual curvature design realises the geometric majority of the potential improvements. For example, at $D=$\SI{300}{\micro\metre}, the dual curvature design offers more than 4-fold of the 7-fold improvement of retroreflective optimisation, and at $D=$\SI{500}{\micro\metre}, 8-fold of the maximum 18-fold improvement. The results for the few-parameter mirrors contain numerical noise as the optimisation is not guaranteed to find a sharp global optimum in the parameter space. However, as exemplified in the sample surfaces of Fig.~\ref{fig: designed surfaces} and discussed in the next section, it is typically preferable (and possible) to find broad peaks in parameter space that have near-optimum performance than sharp peaks with globally-optimum performance.

The performance enhancements of mirror shaping are further contextualised by comparing to concave-concave cavities in aligned and misaligned scenarios (Fig.~\ref{fig: master scan combined}b). Due to the large mode angles involved for misaligned cavities near concentric geometries, we estimate their $\cintdirect$ through power-clipping~\cite{Hunger:10} rather than mode mixing (specific approach detailed in Methods). We see that mirror shaping (exemplified by the retroreflective optimisation case) can bridge some or most of the $\cintdirect$ gap between the plano-concave and concave-concave cavities with spherical mirrors, opening the numerical aperture scaling (Eq.~(\ref{eq: cint concave concave large diameter limit})) to plano-concave geometries. For the largest diameters, optimised plano-concave cavities can even outperform concave-concave cavities that have been designed to tolerate transverse misalignment.

\section*{Sensitivity to parameters and angular misalignment}
\subsection*{Parameter sensitivity}
\begin{figure}[ht!]
\centering\includegraphics[width=15cm]{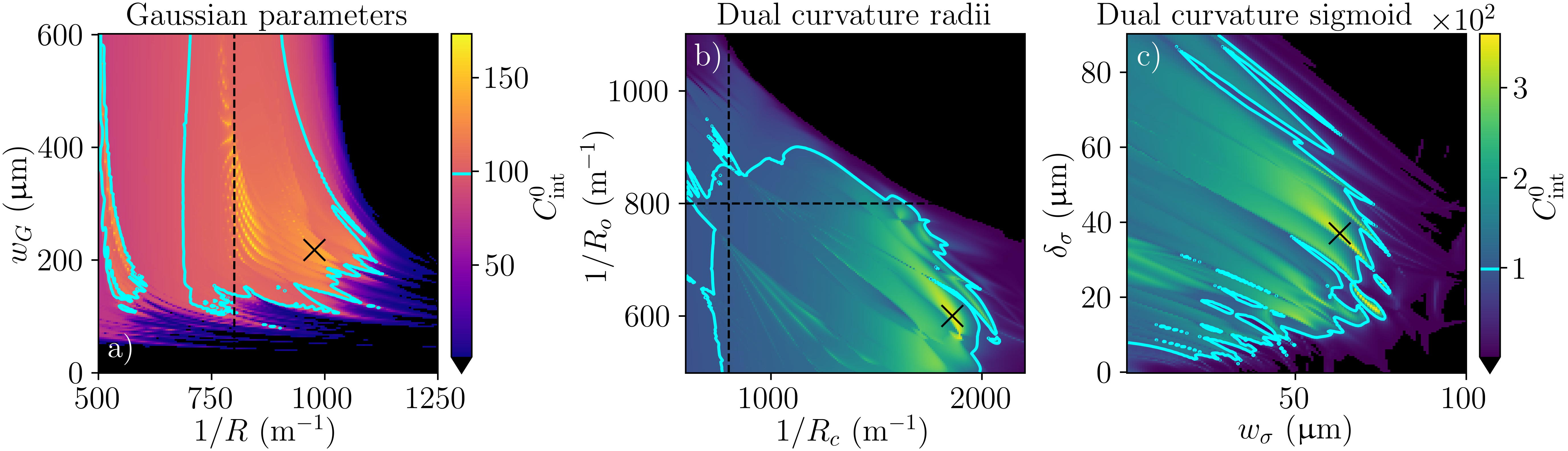}
\caption{Sensitivity of few-parameter mirrors to errors in their shaping parameters for the example cavity geometry ($L=$~\SI{1}{\milli\metre}, $\lambda=$~\SI{1033}{\nano\metre}, $D=$~\SI{300}{\micro\metre}, $\lnotclip=$~20ppm). a) Sensitivity of $\cintdirect$ to the parameters of the Gaussian mirror, where the black cross marks the example Gaussian mirror in Fig.~\ref{fig: designed surfaces}. b-c) Sensitivity of $\cintdirect$ to the b) radial and c) sigmoid parameters of the dual curvature mirror. The black crosses collectively mark the example dual curvature mirror in Fig.~\ref{fig: designed surfaces}, with the black cross in b) marking the static radial parameters for panel c), and vice versa. For all figures, the dotted black line marks $\idealpc{R}$, and the cyan contour $\cintdirectpc$ (Eq.~(\ref{eq: cint pc limit})).}
\label{fig: parameter sensitivity}
\end{figure}

We have seen that a few-parameter mirror can offer the geometric majority of the possible improvement in $\cint$ while avoiding the complex surface shape and extreme length sensitivity of a fully optimised design. To make sure the expected improvement is achieved, the mirror design should not be highly sensitive to fabrication errors that cause changes in the mirror parameters. Examples of mirror parameter sensitivity are shown in Fig.~\ref{fig: parameter sensitivity}. For the Gaussian mirror (Fig.~\ref{fig: parameter sensitivity}a), there is a broad region where $\cintdirect>\cintdirectpc$, but points with the very highest $\cintdirect$ are also very sensitive to errors in fabrication. However, for the dual curvature mirror (Fig.~\ref{fig: parameter sensitivity}b-c), the performance fluctuations appear less strong. In a real use case, the optimum practical design will compromise between performance and sensitivity to fabrication parameters, with the importance of parameter sensitivity informed by the repeatability of the manufacturing process.

\subsection*{Angular misalignment}
The plano-concave geometry is insensitive to the transverse alignment of the mirrors by symmetry (assuming a large planar mirror), but it is sensitive to the angular misalignment (i.e. imperfect orientation of the mirrors with respect to each other). Optical microcavities often require precise angular alignment~\cite{Pallmann:23, Abdelatief:25}, and it is therefore important to know the sensitivity of plano-concave cavities with non-spherical mirrors to determine the passive or activate angular alignment performance required for their operation.

\begin{figure}[ht!]
\centering\includegraphics[width=13cm]{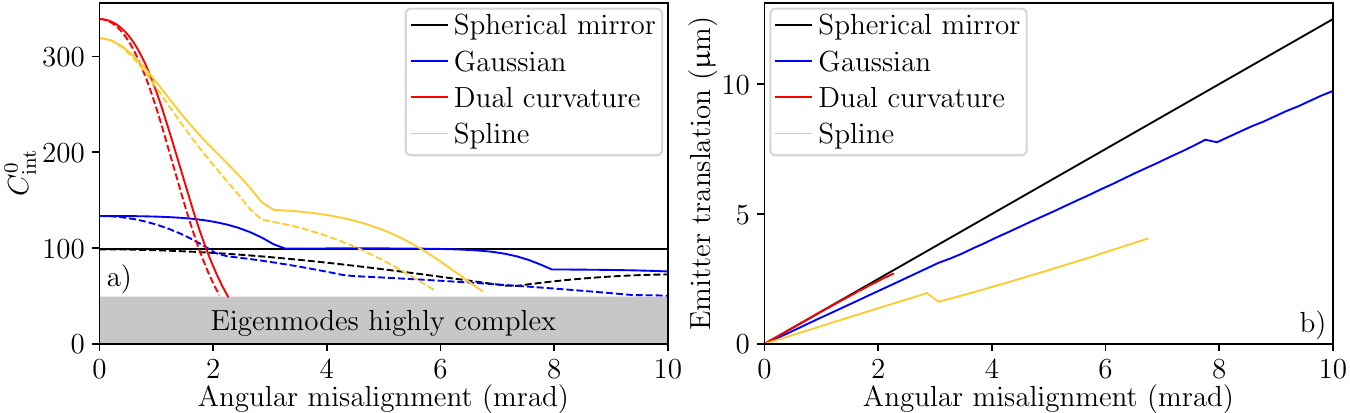}
\caption{Properties of plano-concave cavities, with spherical and few-parameter mirrors, as a function of angular (orientation) misalignment. a) Reduction in $\cintdirect$ for the few-parameter mirror cavities of Fig.~\ref{fig: designed surfaces} ($L=$~\SI{1}{\milli\metre}, $\lambda=$~\SI{1033}{\nano\metre}, $D=$~\SI{300}{\micro\metre}, $\lnotclip=$~20ppm) in the case that (solid) the emitter is translated transversely to track the cavity mode maximum and (dashed) the emitter remains on the axis between the centers of the mirrors. Data not shown for cavities with a $\cintdirect < 0.5 \times\cintdirectpc$ as these cavity eigenmodes are typically completely unlike the eigenmodes in the aligned case, so coupling to them may not be practical. b) The required translation in the case that the emitter tracks the cavity mode maximum.}
\label{fig: angular sensitivity}
\end{figure}

The angular alignment sensitivity of plano-concave cavities with the few-parameter shaped mirrors of Fig.~\ref{fig: designed surfaces} is shown in Fig.~\ref{fig: angular sensitivity} (see Methods section for simulation procedure), where we analyse both the case where the emitter is translated transversely to track the cavity mode maximum, which could apply when the experimental setup is first aligned, and where it is not, which is more relevant for drifts and fluctuations. The cavity with the spherical mirror has very low sensitivity to angular misalignment (Fig.~\ref{fig: angular sensitivity}a), especially if the emitter can be translated (Fig.~\ref{fig: angular sensitivity}b). The non-spherical mirror shapes display higher, although not unreasonable, sensitivity to angular alignment. The reason for the increased sensitivity is the mirror curvature; the curvature of the spherical mirror is constant, so under angular misalignment the mode simply translates (by $\Delta x = R\Delta\theta$, where $\Delta\theta$ is the angular misalignment) to the part of the mirror that is normal to the planar mirror. The mode remains far from the mirror edge precisely due to the space on the non-planar mirror that is not exploited when using spherical mirrors. In contrast, the curvature of the non-spherical mirror is not constant and the mode is larger on the mirror (see Fig.~\ref{fig: designed surfaces}b). Thus, when the mirror is tilted, the mode deforms, affecting the peak mode intensity in the emitter plane and increasing the proportion of the mode that leaks outside the edge of the cavity.

We have seen that using non-spherical cavity mirrors tends to increase the sensitivity of plano-concave cavities to length change (Fig.~\ref{fig: designed surfaces}), angular misalignment (Fig.~\ref{fig: angular sensitivity}), and shape parameter values (Fig.~\ref{fig: parameter sensitivity}). However, that increased sensitivity is only necessarily a problem if the relevant feature in the setup or manufacture cannot be controlled to the required level. Also, the mirror shapes depicted have not been optimised for their robustness, so there are likely to be designs that offer higher tolerance to, for example, angular misalignment at equivalent performance levels to the examples discussed.

\section*{Suggested use cases for plano-concave mirror shaping}
Having found that mirror shaping can improve the performance of plano-concave cavities, we now discuss how to determine whether shaped plano-concave cavities might be useful given a set of geometrical limitations ($L$, $D$, $\lambda$, and $\lnotclip$). This discussion assumes there is a target value of $\cintdirect$ (henceforth $\cintdirecttarget$), above which the cavity performance will be deemed acceptable for the application. Calculating $\cintdirectpc$ (Eq.~(\ref{eq: cint pc limit})), $\cintdirectcclimit$ (Eq.~(\ref{eq: cint concave concave large diameter limit})), and the retroreflective optimised $\cintdirect$ (henceforth $\cintdirectretopt$) establishes four regimes. 

\begin{itemize}
    \item $\cintdirecttarget<\cintdirectpc$: The target performance is achievable with a plano-concave cavity with spherical mirrors.
    \item $\cintdirectpc<\cintdirecttarget<\cintdirectretopt$: Mirror shaping can serve to achieve the target performance in a plano-concave design.
    \item $\cintdirectretopt < \cintdirecttarget < \cintdirectcclimit$: The target performance is likely possible with a concave-concave design, but not with a plano-concave cavity.
    \item $\cintdirecttarget > \cintdirectcclimit$: The target performance is not possible with a concave-concave cavity with spherical mirrors, even if perfectly aligned.
\end{itemize}

If $\cintdirecttarget$ is in the regime where shaped plano-concave cavities may find use, one should weigh their advantages (transverse alignment insensitivity, only one non-planar mirror) against those of a spherical mirror concave-concave design (no need to fabricate non-spherical mirrors). If the plano-concave design is preferred, the length scan, fabrication sensitivity, and angular misalignment sensitivity should be checked to ensure practical tolerances.

\section*{Conclusion}
We have shown that mirror shaping can overcome the key disadvantage of plano-concave cavities with spherical mirrors: The limited coupling strength to a central emitter. We find even simple, manufacturable mirror profiles achieve the majority of the maximum possible performance increases, yield cavities with modest length sensitivity, and retain the misalignment tolerance and simplicity of fabricating only one non-planar mirror inherent to the plano-concave geometry. The potential order-of-magnitude cooperativity improvements could dramatically improve the rate and fidelity of various quantum technology protocols, such as single photon production or qubit state readout. We further discuss how to determine, during experimental design, whether these shaped plano-concave cavities offer a compelling alternative to traditional concave-concave configurations, and anticipate that this work will encourage their use to realise more stable, scalable, and high-performing emitter-cavity systems across quantum technology applications.

\section*{Methods}

\subsection*{Internal cooperativity limit in concave-concave cavities}
As explained in the main text, to maximise $\cint$ of a concave-concave cavity, we should choose $w_0 < \woconfocal$, and decrease $w_0$ until the waist on the mirror causes the clipping loss to exceed $\chi\lnotclip$, where $\chi$ is a small fraction of unity, but not orders of magnitude smaller. This yields a waist on the mirror satisfying
\begin{equation}
    \chi\lnotclip = 2 \exp{-2\left(\frac{D}{2w_m}\right)^2},
    \label{eq: app threshold loss clipping}
\end{equation}
where we have used the power clipping loss expression for both mirrors of diameter $D$ of the cavity~\cite{Hunger:10}. Assuming the high-diameter limit of Eq.~(\ref{eq: waist at mirror}), i.e. $\left(w_0 \ll \woconfocal\right)$, we find $w_m \approx (\woconfocal)^2/w_0$, which leads (through rearrangement of Eq.~(\ref{eq: app threshold loss clipping})) to
\begin{equation}
    w_0 = \frac{2 \left(\woconfocal\right)^2}{D} \sqrt{-\frac{1}{2}\ln{\left(\frac{\chi}{2}\lnotclip\right)}}.
    \label{eq: app loss limited cc w0}
\end{equation}
Applying Eq.~(\ref{eq: gaussian waist internal cooperativity}) from the main text and recognising that the emitter is at the central waist ($w_e=w_0$) leaves
 \begin{equation}
    \cintdirectcclimit = \frac{1}{\left\{-\frac{1}{2}\ln\left({\frac{\chi}{2}\lnotclip}\right)\right\}}\frac{6 D^2}{L^2  \lnotclip}.
\end{equation}

\subsection*{Algorithm and parameters for numerical simulation}
\subsubsection*{Retroreflective optimisation}
The retroreflective optimisation optimises $\cintdirect$ in a basis containing radially symmetric ($m=0$) Laguerre-Gauss modes with radial index $l$ between zero and $l=l_{\mathrm{max}}$. The optimisation varies coefficients of the mode vector that represent amplitudes of the basis mode, but the number of variable coefficients does not have to match the size of the basis. In our simulations, the coefficients of states from $l=0$ to $l=l_{\mathrm{param}}$ are variable, with the remaining coefficients of the mode vector set to zero. The basis for optimisation was the the Gaussian mode family with fundamental mode $\idealpc{\epsilon}$.

The optimisation calculates the expected $\cintdirect$ and its gradient in the space of basis state amplitudes, using an L-BFGS algorithm to optimise~\cite{Hughes:25}. The initial vector of the optimisation is a state with a constant magnitude, alternating sign coefficient applied to the first $s_o$ even-$l$ basis states, with all other coefficients being zero, where $s_o$ is the number of occupied states for which the initial vector maximises $\cintdirect$. This initial vector is chosen because the optimised coefficients usually resemble it.

Once the target mode is calculated, the mirror surface is designed to match its equiphase surface. The phase of the target mode is evaluated at $N_s$ samples across the surface, linearly spaced in radial coordinate. If the sum of any two adjacent phase differences exceeds $\pi$, $N_s$ is deemed insufficient to sample the variation in phase, and the procedure is repeated with $N_s$ doubled. The cavity round trip matrix is calculated from this sampled surface by mode mixing. Diagonalising the round trip matrix yields eigenmodes of which the selected mode has the highest $\cintdirect$. The parameters used for the optimisation are given in Table~\ref{tab: app retroreflective parameters}. 

\begin{table}[h]
    \centering
    
    \begin{tabular}{|c|c|c|c|}
        \hline
        & Fig.~\ref{fig: retroreflective_optimisation_example} & Fig.~\ref{fig: designed surfaces}d & Fig.~\ref{fig: master scan combined} \\
        \hline
        $l_{\mathrm{max}}$ & 120 & 120 & 120 \\
        \hline
        $l_{\mathrm{param}}$ & 79 & 79 & 79 \\
        \hline
        $N_s$ & 5000 & 5000  & 4000 \\
        \hline
    \end{tabular}
    \caption{Parameters used for the retroreflective optimisations conducted in the manuscript. The columns indicate the figure in which the data is presented.}
    \label{tab: app retroreflective parameters}
\end{table}

\subsubsection*{Designed surface optimisation}
We also calculate the modes of cavities with three specific surface types: Gaussian, dual curvature, and a three-point spline. These mode mixing calculations involved basis states from $l=0$ to $l=l_{\mathrm{max}}$, with integrals conducted numerically over $N_s$ linearly-spaced radial samples. The parameters used are summarised in Table~\ref{tab: app specific surface parameters}. The surface parameters were optimised in Fig.~\ref{fig: master scan combined} using the global `differential evolution' genetic algorithm~\cite{Storn:97} with 100 iterations over a population size of 18.

\begin{table}[h]
    \centering
    
    \begin{tabular}{|c|c|c|c|c|}
        \hline
        & Fig.~\ref{fig: designed surfaces}& Fig.~\ref{fig: designed surfaces}f & Fig.~\ref{fig: master scan combined} & Fig.~\ref{fig: parameter sensitivity} \\
        \hline
        $l_{\mathrm{max}}$ & 40 & 120 & 40 & 40 \\
        \hline
        $N_s$ & 4000 & 5000 & 4000 & 4000 \\
        \hline
    \end{tabular}
    \caption{Parameters used for mode mixing simulations of the few-parameter surface types (split by the figure in which the data appears).}
    \label{tab: app specific surface parameters}
\end{table}

\subsubsection*{Simulation of angular alignment sensitivity}
We use mode mixing to simulate the angular alignment sensitivity of plano-concave cavities with shaped mirrors (see Fig.~\ref{fig: angular sensitivity}). As angular misalignment breaks continuous rotational symmetry about the cavity axis, these simulations do not use the Laguerre-Gauss basis restricted to radially symmetric modes (as for all other mode-mixing simulations in the manuscript) but instead use the Hermite-Gauss basis. These basis states are indexed independently in the $x$ and $y$ directions. Without loss of generality, we always rotate the mirror about its $y$-oriented axis, which preserves the $y\rightarrow-y$ symmetry of the cavity, and therefore we only use modes with even parity (and thus even index) in the $y$-direction. 

The mode mixing method treats mirrors as phase surfaces occupying a single plane rather than physical objects in space. Therefore, to rotate a mirror with initial surface $z(x,y)$ by an angle $\theta$ about its $y$-directed axis, we simply simulate the modified profile
\begin{equation}
    z'(x,y) = z(x,y) + \theta x,
\end{equation}
which uses the small angle approximation as all angular misalignments considered are small.

Once the round trip matrix of the misaligned cavity is calculated, we must select the appropriate eigenmode. Here, we take the cases that the emitter could not be translated, and that the emitter could be freely translated in $x$ to track displacements to the cavity mode. In the case that the emitter is not translated, the mode that maximises $\cintdirect$ at the original emitter position is chosen. In the case that the emitter is translated, the translation is optimised numerically with the mode that maximises $\cintdirect$ chosen for each translation; We optimise using a local `Nelder-Mead' method with the initial value being the expected translation from geometric theory with the best spherical mirror. Our mode basis contained the first 60 $x$-indices and the first 25 even $y$-indices. The grid of the numerical integration contained $100^2$ equal areas over a total square of side length equal to the mirror diameter. Areas lying outside of the mirror bounds were given zero weight in the numerical integral.

\subsubsection*{Spherical surface optimisation}
The data in Fig.~\ref{fig: master scan combined} presents the optimum $\cintdirect$ available for cavities with spherical non-planar mirrors in the plano-concave and transversely aligned concave-concave cases as a function of the mirror diameter. We optimise a single parameter controlling the radius of the non-planar mirror (discussed later). Because we optimise just one rather than many parameters, we use Bayesian optimisation (the bayesian-optimisation Python package~\cite{Nogueira:14}) instead of differential evolution. The optimisation routine takes $S_i$ initial points uniformly over the range (discussed later), and then optimises over $S_{\mathrm{opt}}$ further samples. In each trial, the cavity mode was calculated through mode mixing, using $N_s$ linearly-spaced samples of a radial integral in a basis using states from $l=0$ to $l=l_{\mathrm{max}}$. For both plano-concave and concave-concave optimisations, we used $l_{\mathrm{max}}=120$, $N_s=4000$, $S_i=50$, and $S_{\mathrm{opt}}=150$.

For plano-concave geometries, the optimised parameter was the inverse radius of curvature of the non-planar mirror, which ranged from $1/(2L)$ to $4/(5L)$. The lower bound minimises the size of the mode on the spherical mirror, and the upper bound, which is $\idealpcr$, produces the smallest Gaussian waist at the emitter. The basis for optimisation was the the Gaussian mode family with fundamental mode $\idealpc{\epsilon}$. For the concave-concave case, the optimised parameter was $w_0$, the central waist of the basis (located at the emitter for this geometry). We did not use the mirror curvature $R$ directly because large changes in mode divergence and $w_0$ are seen approaching the concentric limit of $R=L/2$. The value of $w_0$ was bounded from below by $\lambda$, and above by $\woconfocal$, which minimises the mode size on the mirror. The lower limit of $\lambda$ is used because our approach invokes the paraxial approximation, and focussing to a waist of $\lambda$ would break the validity of this approximation. To confirm the paraxial approximation is appropriate, it was verified post-calculation that the optimum $w_0$ was always significantly above $\lambda$.

\subsection*{Method for calculating clipping losses in misaligned concave-concave cavities}
\label{app: misaligned clipping method}
The results of Fig.~\ref{fig: master scan combined}b require the calculation of the clipping loss for misaligned concave-concave cavities. The mode mixing method is suitable in principle, but for the large mode axis angles possible with wide mirrors, a very large basis size is required. Instead, we use standard cavity theory~\cite{Blows:98} to calculate the expected mode of the cavity with the given mirror curvatures, length, and misalignment, and then use a clipping integral to calculate the power that is lost on reflection. While this method does not produce exactly correct results, it gives answers of the correct order of magnitude~\cite{Hughes:23}, which is sufficient to estimate the achievable performance. 

\begin{figure}[ht!]
\centering\includegraphics[width=7.0cm]{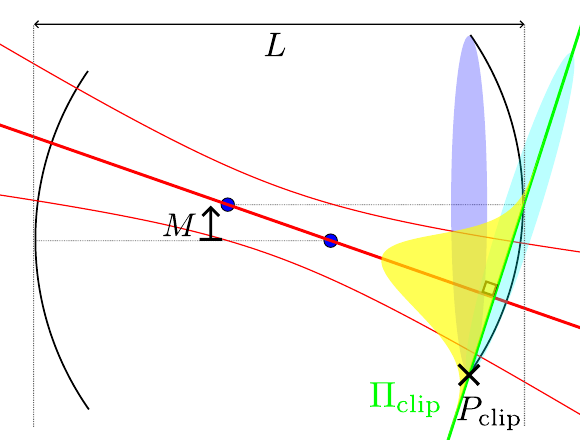}
\caption{Diagram of the clipping loss calculation for a misaligned concave-concave cavity with length $L$ and mirror misalignment $M$ used for results in Fig.~\ref{fig: master scan combined}b. The mode (thick red line, with thin red lines indicating the $1/e$ power waist) predicted by classical theory propagates along the line between the centres (black dots) of the circles that define the two mirrors. The point on the edge of the mirror where the mode has highest intensity is $P_{\mathrm{clip}}$. The clipping plane $\Pi_{\mathrm{clip}}$ is defined as the plane perpendicular to the mode running through $P_\mathrm{clip}$. The power retained in the mode upon reflection is calculated by integrating the intensity profile of the mode in $\Pi_{\mathrm{clip}}$ (yellow Gaussian) over the mirror aperture rotated into this plane (cyan oval).}
\label{fig: app cc clipping}
\end{figure}

Unfortunately, the power clipping integral only produces normalised results if the plane of the power integral is perpendicular to the propagation of the beam. However, the mirror aperture in a misaligned cavity is generally not perpendicular to the mode because the mode axis is tilted. In Fig.~\ref{fig: app cc clipping} we depict our choice of power integral plane as perpendicular to the mode (so that the integral is well-normalised) and passing through the edge on the mirror where the mode has the highest intensity, and therefore clipping is most severe.

To calculate optimum performance values for a given $L$, $D$, and $M$, we scan the curvature $R$ of the mirrors (through a proxy variable explained later), calculating $\cintdirect$ for each $R$, and choosing the best of these results. Instead of scanning $R$ directly, we scan the `aligned central waist' (the $w_0$ the cavity would have if the misalignment were zero) and calculate $R$ from this in order to avoid a parameterisation where the changes in mode properties diverge as $R\rightarrow L/2$ from above.

\subsection*{Data availability}
The datasets generated and/or analysed during the current study are available in the University of Southampton Institutional Repository \href{ https://doi.org/10.5258/SOTON/D3719}{DOI: 10.5258/SOTON/D3719}.
The source code that produced the data is available from the corresponding author (\href{mailto:w.j.hughes@soton.ac.uk}{w.j.hughes@soton.ac.uk}) at reasonable request.

\section*{Acknowledgements}
The authors acknowledge the use of the IRIDIS High Performance Computing Facility, and associated support services at the University of Southampton, in the completion of this work.

\section*{Author contributions statement}

W.J.H and P.H conceived of the study. W.J.H performed the numerical simulations. All authors contributed to writing and reviewing the manuscript. 

\section*{Funding}
W.J.H and P.H were both funded by the UK Engineering and Physical Sciences Research Council Hub for Quantum Computing and Simulation (EP/T001062/1) and Hub for Quantum Computing via Integrated and Interconnected Implementations (EP/Z53318X/1).

\section*{Competing interests} The authors declare no competing interests.

\end{document}